\documentstyle[12pt]{article}
\long\def\comment#1{}
\begin{document}
\title{Aristotle and Gautama on Logic and Physics}
\author{Subhash Kak\\
Louisiana State University\\
Baton Rouge, LA 70803-5901\\}
\maketitle

\subsection*{Introduction}

Physics as part of natural science and philosophy
explicitly or implicitly uses logic, and from this point of 
view all cultures that studied nature [e.g. 1-4] must have had 
logic.  But 
the question of the origins of logic as a formal discipline is of 
special interest to the historian of physics since it represents a turning
inward to examine the very nature of reasoning and the relationship between
thought and reality. 
In the West, 
Aristotle (384-322 BCE) is generally credited with the formalization of 
the tradition of logic and also with the development of early 
physics.
In India, the \d{R}gveda itself in the hymn 10.129 suggests the beginnings
of the representation of reality
in terms of various logical divisions
that were later represented formally
as the four circles of {\it catu\d{s}ko\d{t}i}: ``A", 
``not A", ``A and not A", and ``not A and not not A'' [5].
Causality as the basis of change was enshrined in the early philosophical system of
the S\={a}\d{m}khya.
According to Pur\={a}\d{n}ic accounts, Medh\={a}tithi
Gautama and Ak\d{s}ap\={a}da Gautama (or Gotama), 
which are perhaps two variant names for
the same author of the early formal text on Indian logic, belonged
to about 550 BCE.

The Greek and the Indian traditions seem 
to provide the earliest formal representations of logic, and in this article we 
ask if
they influenced each other.
We are also interested in the scope of early logic, since this gives
an idea to us of the way early thinkers thought about nature and change.
We will show that Greek and Indian logical traditions have much that is
distinctive and unique.

Philosophy and physics were considered part of the same intellectual 
enterprise until comparatively recent times.
Thomas
McEvilley's {\it The Shape of Ancient Thought} does an excellent
comparative analysis of Greek and Indian philosophy [6], stressing how 
there existed much interaction between the two cultural areas in very early
times, but he argued that they evolved independently.
Some scholars believe that the five part syllogism of 
Indian logic was derived from the three-part Aristotelian logic. On the other 
hand, there is an old tradition preserved by the Greeks and the Persians which 
presents the opposite view. According to it, 
Alexander was the intermediary who brought Indian logic to the Greeks and it
was under this influence that the later Greek tradition emerged.

In this article, I review the evidence afresh and conclude that although
the Greeks may have been aware of 
Indian logic, there is no reason to suppose that it influenced their own
development of the subject in any fundamental 
way. I argue that the five-part Indian
syllogism is likely to have been an earlier invention than Aristotle's work.
These conclusions
can assist us in finding the chronological framework for the development of logic within India.

\subsection*{Background}

My interest in the question of the connections between Indian and 
Greek logic was triggered
several years ago by  a letter from Andrei Heilper of the IBM 
Research Laboratory in Haifa, Israel.
Heilper felt that Western 
logic might be indebted to Indian logic but this feeling was based on 
secondary evidence, and he wanted to know if I could help him reach a more 
definitive conclusion.

Heilper's interest in Indian logic was born out of a passage in the book
{\it The World as Will and Representation} [7] by the German 
philosopher Arthur Schopenhauer (1788-1860), who was commenting on the reference 
by the Indologist 
William Jones (1746-1794) on this matter.$^1$

William Jones has the following account on the question of the relationship
between Indian and Greek logic (it appears
in the 11th discourse):$^2$

\begin{quote}
Here I cannot refrain from introducing a singular tradition, which
prevailed, according to the well-informed  author of the ``Dabistan'', in the 
Panjab
and in several Persian provinces, that, ``among other Indian curiosities,  which
CALLISTHENES transmitted to his uncle, was a technical system of logick, 
which the Brahmens had  communicated  to the inquisitive Greek,'' and which the
Mohammedan writer supposes to have been the  groundwork of the famous
Aristotelean method:  if this be true, it is one of the most interesting facts,
that I have met with in Asia; and if it be false, it is very extraordinary, 
that such a
story should have been fabricated either by  the candid MOHSANI Fani; or by the
simple Parsis Pundits, with whom he had conversed; but, not having had leisure
to study the Nyaya Sastra, I can only assure you, that I have frequently seen
perfect syllogisms in the philosophical writings of the Brahmens, and have 
often
heard them used in their verbal controversies. ...
\end{quote}

Kallisthenes (370-327 BCE), a relative of the philosopher Aristotle,
was the court 
historian to Alexander who was a member of the campaign. He 
was executed by Alexander in 327 BCE.
There exists credible evidence that Kallisthenes was asked by Aristotle to 
bring
texts to Greece. 
Since his bringing back of the astronomical observations of the
Babylonians is attested by several sources, it is reasonable to assume that the story
about his having brought back Indian logic is also credible.
But this cannot be taken to mean that the texts of Indian logic directly or
indirectly influenced Aristotle.

The relevant passage from the {\it Dabistan-i Madhahib} (School
of Sects), a 17th century
text by Mohsan Fani, a Kashmiri scholar of Persian ancestry
who lived during 1615 - 1670 (?) [8, pages 270-273]:

\begin{quote}
{\it Tark sastra} is the science of dialectics; ...
These are the sixteen parts of the Tarka.  The followers of this doctrine judge
and affirm that, as this world is created, there must be a Creator; the 
{\it mukt}, or ``emancipation,'' in their opinion means striving to approach the origin
of beings, not uniting like the warp and the web, the threads of which, 
although
near, are nevertheless separate from each other.  This was related to me by the
Imam Arastu [Aristotle], who was a chief of the learned and said to me that 
he had derived it from an old treatise upon logic, the precepts of which 
were without
explanation, and to have bestowed on it that arrangement under which it now
exists among the learned:  he meant, probably, that the maxims are the same as
those extracted from the Tarka.  The same doctrine was taught in Greece; in
confirmation of this, the Persians say, that the science of logic which was
diffused among them was, with other sciences, translated into the language of
Yonia and Rumi, by order of king Secander [Alexander], the worshiper of 
science, in the time of his conquest, and sent to Rumi.
\end{quote}

The similarity in the reasoning of pre-Christian Greece and India was noted 
repeatedly by al-Biruni (1030 CE) in  his book on India [9].
The Indian system takes the material processes to be governed by laws in the
ancient philosophical framework of S\={a}\d{m}khya, which takes the
evolution and change in the world to be entirely materialistic while acknowledging
the existence of consciousness as a separate category.

\subsection*{Aristotle's logic}

In the West, Aristotle's 
theory of the syllogism has had enormous influence. At 
one time in Greece, Stoic logic was more influential until Aristotles 
ideas became dominant, and they were subsequently adopted by the Arabic and 
the Latin medieval traditions. The commentators grouped Aristotle's works 
on logic under the title {\it Organon} (Instrument), which comprised of:

\begin{enumerate}
    \item Categories
    \item On Interpretation
    \item Prior Analytics
    \item Posterior Analytics
    \item Topics
    \item On Sophistical Refutations
\end{enumerate}

The central notion in Aristotle's logic is that of deduction
involving premise  of the argument, and the conclusion.
It also recognizes
induction, which is an argument from the particular to the 
universal.

In Aristotle's syllogism, 
the primary premise 
is always universal, and it may be positive or negative. The secondary premise 
may also be universal or particular so that from these premises it is 
possible to deduce a valid conclusion.
An example is : ``All men are mortal; Socrates is a man; 
therefore, Socrates is mortal.'' Another: 

\begin{itemize}
    \item Everything that lives, moves ({\it primary premise})
    \item No mountain moves ({\it secondary premise})
    \item No mountain lives ({\it conclusion})
\end{itemize}

Aristotle supposed that this scheme accurately represents
the true nature of thought.
If we take thought,
language, and reality to be isomorphic, consideration of our reasoning will
help us understand reality.

In {\it Categories}, Aristotle
makes a distinction among three ways in
which the meaning of different uses of a predicate may be related to each
other: {\it homonymy, synonymy, and paronymy} (``equivocal,'' ``univocal,'' and
``derivative''). 
For any such use, he proposed descriptions in ten attributes:
substance, 
quantity, quality, relative, where, when, being
in a position, having, acting on, and being affected by.
The most important of these is substance, which is the individual thing itself;
secondary substances include the species and genera to
which the individual thing belongs. 
The other categories distinguish
this individual substance from others of the
same kind.

In {\it Interpretation}, Aristotle
considers the use of predicates in combination with
subjects to form
propositions or assertions, each of which is either true or false. But he 
recognizes that certain difficulties arise 
when speaking of the future.
He suggests that it is
necessary that either tomorrow's event will occur or it will not, but it is
neither necessary that it will occur nor necessary that it will not occur.

In {\it Prior Analytics}, Aristotle used mathematics as a model to show that
knowledge must be derived from what is already known. The process of
reasoning by syllogism formalizes the 
the deduction of new truths from established principles. 
He offered a
distinction between the non-living and the living in terms of things that 
move only when moved by something else and those that are capable of
moving themselves. 
He also 
distinguished between the basic material and the form and purpose which jointly
define the 
individual thing. 

He suggested four different explanatory principles or causes in his {\it Physics}:
(i) {\it the material cause}
 is the basic substance out of which the thing is made (like the 
building materials for a house);
(ii) {\it the formal cause}
is the pattern in conformity with which these materials
are assembled (like the builder's plan of the house); 
(iii) {\it the efficient cause} is the agent or force immediately responsible for
the production of the thing (builders of the house); and (iv) {\it the final cause}
is the end
purpose for which a thing exists (shelter to the resident of the house).
Aristotle believed that the four causes are essential in the existence and
nature of all things.

The four causes apply clearly to machines built by man.
As for things that appear to arise by pure chance,
Aristotle believed that there must likewise be four causes of which we
may not be unaware.

Aristotle's logic has been the basis of theology in the West.
Modern science rejects the notion of final causes. Creationism and theories of
Intelligent Design are an attempt to bring in final causes into biology.

\subsection*{The Indian tradition}

Logic ({\it \={a}nv\={\i}k\d{s}ik\={\i}, ny\={a}ya}, or {\it tarka} in Sanskrit) is 
one of the six 
classical schools of Indian 
philosophy. 
These six schools are the different complementary
perspectives on reality, that may be visualized as the views from the windows
in the six walls of a cube within which the subject is enclosed. 
The base is the system of the traditional
rites and ceremonies (P\={u}rva M\={\i}m\={a}\d{m}s\={a}), and the ceiling is 
reality that includes the objective world and
the subject (Uttara M\={\i}m\={a}\d{m}s\={a} or Ved\={a}nta); one side is 
analysis of linguistic particles (Ny\={a}ya), with the
opposite side being the analysis of material particles (Vai\'{s}e\d{s}ika); another side
is enumerative categories in evolution at the cosmic and individual levels (S\={a}\d{m}khya),
with the opposite side representing the synthesis of the material and cognitive systems
in the experiencing individual (Yoga).
Clearly, the use of systematic view of nature had been taken to a very advanced level.

Logic is described in Kau\d{t}ilya's Artha\'{s}\={a}stra (c. 350 BCE)
as an independent field of inquiry {\it \={a}nv\={\i}k\d{s}ik\={\i}} [10].
The epic Mah\={a}bh\={a}rata, which is most likely prior to 500 BCE because it is not
aware of Buddhism in its long descriptions of religion [11],  
declares
(Mah\={a}bh\={a}rata 12.173.45) that {\it \={a}nv\={\i}k\d{s}ik\={\i}} is equivalent
to the discipline of {\it tarka}. Clearly, there were several equivalent terms in use in
India for logic in 500 BCE. 

The canonical text on the Ny\={a}ya is the Ny\={a}ya S\={u}tra of Ak\d{s}ap\={a}da
Gautama.
The most important early 
commentary on this text is the Ny\={a}ya Bh\={a}\d{s}ya of V\={a}tsy\={a}yana which
is estimated to belong to 5th century CE.
Satisa Chandra Vidyabhusana, the well-regarded authority
on Indian logic, assigned
Ak\d{s}ap\={a}da Gautama the date of approximately 550 BCE.  He based this on the 
reference in the K\={a}\d{n}va recension of the \'{S}atapatha Br\={a}hma\d{n}a, 
in which 
Gautama (or Gotama) is shown to be contemporary of J\={a}tukar\d{n}ya Vy\={a}sa, who 
was a student of \={A}sur\={a}ya\d{n}a.
This and other evidence from the G\d{r}hya S\={u}tras, the V\={a}yu 
Pur\={a}\d{n}a, and the Buddhist
Sanskrit text the Divy\={a}vad\={a}na  is summarized in the 
introduction of Vidyabhusana's edition of the
Ny\={a}ya S\={u}tra [12].

In his ``History of Indian Logic'', Vidyabhusana modified his views [13] under
the influence of the then current ideas of history of science and the now-discredited 
Aryan
invasion theory. 
He now argued that the texts
speak of two Gautamas who are both associated with logic.
He declared that Medh\={a}tithi Gautama (was) the founder of 
{\it \={a}nv\={\i}k\d{s}ik\={\i}} (circ 550 BCE), and Ak\d{s}ap\={a}da Gautama came much later, perhaps 150 
CE or so.

Indian science and chronology has come in for a major revision in recent years.
Archaeologists have found no evidence for any large migration into India subsequent
to 4500 BCE,
and found that Indian art, iconography,
social organization, and cultural motifs
can be traced to a tradition that began in 7000 BCE [14-17]. 
There is also new textual analysis that pushes back the origins
of Indian astronomy and mathematics considerably [18-21].
We see Indian scientific ideas develop in a systematic manner
over a period of several centuries going back to the second millennium
BCE, if not earlier [22-29].
This development is an enlargement in different fields
of the recursive system of Vedic cosmology [5].
We also know that
ancient India and West Asia had considerable interaction much before the
time of Alexander [30].

Parenthetically, we note that genetic evidence
related to mitochondrial DNA and the Y chromosome has allowed the reconstruction
of the movements of ancient peoples. According to the highly regarded synthesis
of this evidence by Stephen Oppenheimer [31], India 
was populated by people who left Africa about 90,000 years ago, and
that in India over the next tens of thousands of years, both the Mongoloid and
the Caucasoid types evolved, migrating to the northeast and 
northwest regions about 50,000 years ago.

The earlier chronology of Indian texts was coloured first by the then
popular Biblical chronology that took the origin of mankind to go back to only
4004 BCE, 
and later by the 
theory of Aryan invasions for which archaeologists
have found no support [15-17]. According to Stephen Oppenheimer, genetic evidence 
also goes counter to the Aryan invasion theory.
One must, therefore, question Vidyabhusana's 
revision of the chronology of Indian
logic, especially in light of the new understanding
of the literary
tradition and the evidence related to the presence of Indian kingdoms in West Asia
in the second millennium BCE that provides a late material basis for the cultural 
continuity 
between India and the West.

\subsection*{Gautama's Ny\={a}ya S\={u}tra}

The Ny\={a}ya also calls itself
{\it pram\={a}\d{n}a \'{s}\={a}stra}, or the science of correct 
knowledge. Knowing is based on four conditions: (i) The subject or the 
{\it pramat\d{r}}; (ii) The object or the {\it prameya} 
to which the process of cognition is 
directed; (iii) The cognition or the {\it pramiti}; and (iv) the nature of knowledge, 
or the {\it pram\={a}\d{n}a.}
The four pram\={a}\d{n}as 
through which correct knowledge is acquired are:  {\it pratyak\d{s}a} or direct 
perception, {\it anum\={a}na}
 or inference, {\it upam\={a}na} or analogy, and 
{\it \'{s}abda} or verbal 
testimony.

The function of definition in the Ny\={a}ya is to state essential nature 
({\it svar\={u}pa}) that distinguishes the object from others. Three fallacies of 
definition are described: {\it ativy\={a}pti}, or the definition being too broad as 
in defining a cow as a horned animal; {\it avy\={a}pti}, or too narrow; and 
{\it asambhava}, or impossible.

Gautama mentions that four factors are involved in direct perception: the 
senses ({\it indriyas}) , their objects  ({\it artha}), the contact of the senses and 
the objects ({\it sannikar\d{s}a}), 
and the cognition produced by this contact ({\it j\~{n}\={a}na}). 
The five sense organs, eye, ear, nose, tongue, and skin have the five 
elements light, ether, earth, water, and air as their field, with 
corresponding qualities of color, sound, smell, taste and touch.

{\it Manas} or mind mediates between the self and the senses. When the manas is 
in contact with one sense-organ, it cannot be so with another. It is 
therefore said to be atomic in dimension. It is due to the nature of 
the mind that our experiences are essentially linear, although quick 
succession of impressions may give the appearance of simultaneity.
 
Objects have qualities which do not have existence of their own. The color 
and class associated with an object are secondary to the substance. 
According to Gautama, direct perception is inexpressible. Things are not 
perceived as bearing a name.  The conception of an object on hearing a name 
is not direct perception but verbal cognition.

Not all perceptions are valid. Normal perception is subject to the
existence of (i) the object of perception, (ii) the external medium such as 
light in the case of seeing, (iii) the sense-organ, (iv) the mind, without which 
the sense-organs cannot come in conjunction with their objects, and (v) the 
self. If any of these should function improperly, the perception would be 
erroneous. The causes of illusion may be {\it do\d{s}a} (defect in the sense-organ), 
{\it samprayoga} (presentation of only part of an object), or {\it sa\d{m}sk\={a}ra}
(habit based on irrelevant recollection).

{\it Anum\={a}na} (inference) is knowledge from the perceived about the
unperceived. The relation between the two may be of three kind: the element 
to be inferred may be the cause or the effect of the element perceived, or 
the two may be the joint effects of something else.

The Ny\={a}ya syllogism is expressed in five parts: 
\begin{enumerate}
\item {\it pratij\~{n}\={a}}, or the
proposition: the house is on fire; 

\item {\it hetu}, or the reason: the  smoke; 

\item 
{\it d\d{r}\d{s}\d{t}\={a}nta} the example:  
fire is accompanied by smoke, as in the kitchen; 

\item {\it upanaya}, the application: as in kitchen so for the house; 

\item {\it nigamana}, 
the conclusion: therefore, the house is on fire.  
\end{enumerate}

This recognizes that the 
inference derives from the knowledge of the universal relation ({\it vy\={a}pti}) and 
its application to the specific case ({\it pak\d{s}adharmat\={a}}). There can be no 
inference unless there is expectation ({\it \={a}k\={a}\.{n}ksh\={a}}) about the 
hypothesis which 
is expressed in terms of the proposition.

The minor premise ({\it pak\d{s}adharmat\={a}}) is a consequence of perception, whereas 
the major premise ({\it vy\={a}pti}) results from induction. But the universal 
proposition cannot be arrived at by reasoning alone. Frequency of the 
observation increases the probability of the universal, but does not make 
it certain. Gange\'{s}a, a later logician, suggested that the apprehension of 
the universal requires {\it alaukika pratyak\d{s}a} (or nonsensory apprehension).

The Ny\={a}ya system lays stress on antecedence in its view of causality. But 
both cause and effect are viewed as passing events. Cause has no meaning 
apart from change; when analyzed, it leads to a chain that continues 
without end. Causality is useful within the limits of experience, but it 
cannot be regarded as of absolute validity. Causality is only a form of 
experience.

The advancement of knowledge is from {\it upam\={a}na}, or comparison, with something 
else already well-known. The leads us back to induction through {\it alaukika 
pratyak\d{s}a} as the basis of the understanding.

{\it \'{S}abda}, or verbal testimony, is a chief source of knowledge. The
meaning of words is by convention. The word might mean an individual, a 
form, or a type, or all three. A sentence, as a collection of words, is 
cognized from the trace ({\it sa\d{m}sk\={a}ra}) left at the end of the sentence. 
Knowledge is divided into cognitions which are not reproductions of former 
states of consciousness ({\it anubhava}) and those which are 
recollections ({\it sm\d{r}ti}). 

The Ny\={a}ya speaks of errors and fallacies arising by interfering with the 
process of correct reasoning. The Ny\={a}ya attacks the Buddhist idea that no 
knowledge is certain by pointing out that this statement itself contradicts 
the claim by its certainty.  Whether cognitions apply to reality must be 
checked by determining if they lead to successful action. 
{\it Pram\={a}}, or valid 
knowledge, leads to successful action unlike erroneous knowledge ({\it vipary\={a}ya}).

\subsection*{More on Gautama's logic and physics}
Gautama's propositions assume a dichotomy between object and subject. The objective
world is open to logical analysis since it maps to linguistic categories; the
subjective world can suffer from invalid perception for a variety of reasons. This is
consistent with the Vedic view that the although the inner world maps the outer, the mind
can be clouded by habits or wrong deductions owing to incorrect assumptions.

The S\={a}\d{m}khya, attributed to the legendary
rishi Kapila, is the first philosophy of science that arose centuries before
the Buddha and it is the background to be considered when speaking of Indian logic.
In S\={a}\d{m}khya, evolution occurs due to changing balance and proportion both in the
objective and the subjective worlds.
The three {\it gu\d{n}as} or fundamental modalities are {\it sattva, tamas} and
{\it rajas}, and they operate both at the large scale as well as
in quick transformation.
The normative ``thing'' behind this ceaseless change is the
witness, or self, who is viewed in the singular for the entire universe.

At the objective level, {\it tamas} is inertia, {\it rajas} is action or
transformation and {\it sattva} is the relative balance or equilibrium between {\it tamas}
and {\it rajas}.
The interplay between the three sets up oscillations in the objective and
the mental levels. In Yoga, the objective is to achieve
the cessation of the fluctuations of the mind. 

Consciousness or pure awareness is by definition not an object and
therefore it does not have attributes. It must for the same reason be
beyond the categories of the living or dead. It must be beyond inertia, or change or
fluctuations. 
It is extraordinary that in this
analysis the qualities that are associated with objects become describable
by an internal order.

The {\it gu\d{n}as} do not admit of any further breakdown.
This defines a position that is deeper and different from that of Aristotelian
physics.
The three {\it gu\d{n}as} are present in
all objects and we can isolate one only in terms of the momentary strength
of one in relation to the other in a process.
Their fluctuations mark
the universal ``internal clock" of worldly
processes. 

In the S\={a}\d{m}khya, the effect is the cause
in a new form, and this is why the system is also  called {\it pari\d{n}\={a}mav\={a}da}, or
theory of transformation.
Between the cause and effect is a relation of identity-and-difference, that is
identity of stuff but difference of form ({\it bhed\={a}bheda}). 
The method at the basis of the S\={a}\d{m}khya and the Ny\={a}ya S\={u}tra may be seen in the Yoga S\={u}tra as well.
In the Yoga S\={u}tra 3.13 three aspects of change are identified: transformation of a
thing ({\it dharmi}) into a property ({\it dharma}), transformation of a property into a mark 
({\it lak\d{s}a\d{n}a}), and the transformation of a mark into a condition
({\it avasth\={a}}).
This is then the basis of the ``unreasonable effectiveness'' of mathematics 
in the description of the world. 

\subsection*{The form of the Ny\={a}ya syllogism}
The five parts of the Ny\={a}ya syllogism spring from the idea of {\it bandhu} that
is fundamental to Vedic thought. The {\it bandhu} is the equivalence between two
different systems, which ordinarily are the microworld, the macroworld, and the
individual's cognitive system.

The Ny\={a}ya syllogism first sets up the propositional system with its two
components ({\it two parts}) and then identifies another well known system to which the
first is supposed to have a {\it bandhu-}like relationship ({\it third and fourth
parts}). The conclusion ({\it fifth part})
can be made only after the preliminaries have been formally defined. As we see, this 
takes five steps.

The appeal to the {\it bandhu} in the syllogism is to acknowledge the agency of
the subject who can be, without such knowledge, open to invalid perception. One can see
how in systems that do not
accept transcendental reality (such as Aristotle's or Buddhist), a simplification from
the five-part to the three-part syllogism would be most natural.

         It was pointed out in a 1930 article by
the French  Indologist  Rene Gu\'{e}n\'{o}n [32-33], that some later Indian texts  
on logic present 
         two abridged  forms of the five-membered
         syllogism of the Ny\={a}ya S\={u}tra, in  which  either  the  first  three
         or the  last  three   parts  appear
         alone.  The  first version is similar to the 6th century
Ny\={a}yaprave\'{s}a of India (which had much influence on
         Chinese Buddhism and on Jain thought), and the latter
         version resembles the syllogism of Aristotle.

It appears then that the Ny\={a}ya S\={u}tra syllogism was only the most
comprehensive way of establishing the chain of reasoning consistent
with Vedic ideas of {\it bandhu}, and since two 
different simplifications arose out of it, it is likely to
have been the older tradition. 

\subsection*{Concluding remarks}

We find that as {\it \={a}nv\={\i}k\d{s}ik\={\i}} Indian logic goes back to at least 
the 6th century
BCE based on the textual evidence that has been universally accepted. 
The syllogism used in the Ny\={a}ya S\={u}tra has five parts, as against the
three-part syllogism of Aristotle's logic. But we know that simplification
of the five-part logic into one similar to Aristotle's was also known in India.

It seems that although the
 Greek and Persian stories related to Kallisthenes
having brought Indian logic to Greece may have a historical basis, 
 they are not to be taken as the
literal truth. At best, Indian logic provided inspiration in
the sense of the need for a formal text on the subject by the Greeks.
The focus in the two logical traditions is quite different, and either one
is unlikely to have borrowed from the other.

Regarding the transmission of the Indian texts, this could have occurred only
at the tail-end of Alexander's campaign of conquest.
It is quite likely that Aristotle's logical texts were already in place by
that time.
It is more likely that the Greek and
Persian traditions merely acknowledge a different 
system of logic
in a civilization that existed far away.

The Ny\={a}ya S\={u}tra and Aristotle's texts are two different
perspectives that tie in with the cosmologies of the two civilizations.
The Ny\={a}ya S\={u}tra, like other Indian philosophical texts,
maintains the centrality of the subject, whereas in  Aristotle's logic
the emphasis is on the design of the world as a machine. This also corresponds to
the difference in the Indian system which considers the universe to be composed of
five elements, as against the four of Aristotle. The fifth element ({\it \={a}k\={a}\'{s}a})
of the Indian system concerns the field of sentience.

Diogenes Laertius in his {\it Lives of the Eminent Philosophers} [34]
says this of Democritus: ``According to Demetrius in his book on Men of the Same Name and
Antisthenes in his Successions of Philosophers, he [Democritus] traveled into Egypt to
learn geometry from the priests, and he also went into Persia to visit the
Chaldaeans as well as to the Red Sea.  Some say that he associated with the
Gymnosophists in India and went to Aethiopia'' [page 445].
   Judging from the rest of the biographical material in Diogenes' account, a
visit to India is not entirely implausible.  Diogenes represents Democritus
as a ``student of the Magi'' from boyhood, and an enthusiastic and well funded
traveler.
This is evidence of the interaction between India and Greece that goes back
much before Aristotle. The legend  that  Indian logic was taken by Alexander to
Greece is just the acknowledgement that Indians had a fully flowered system
of logic before the time of Aristotle, who himself perhaps only reworked an earlier
Greek tradition.

\subsubsection*{Acknowledgement}
I am thankful to Andre Heilper for asking
me about the relationship between
Greek and Indian logic, and to Linda Johnsen for discussion and drawing my attention to the
book by Diogenes Laertius.

\subsubsection*{Notes}
\begin{enumerate}
\item
``Aristotle collected, arranged, and corrected all that had been
discovered before his time, and brought it to an incomparably greater state of
perfection. If we thus observe how the course of Greek culture had prepared the
way for, and led up to the work of Aristotle, we shall be little inclined to
believe the assertion of the Persian author, quoted by Sir William Jones with
much approval, that Callisthenes found a complete system of logic among the
Indians, and sent it to his uncle Aristotle'' [7].
\item
Quoted in the note by Heilper.
\end{enumerate}

\subsection*{Bibliography}

\begin{description}

\item [1]
S. Kak, ``Indian physics: outline of early history.'' ArXiv: physics/0310001.

\item [2]
A. Pannekoek, {\it A History of Astronomy.} Allen and Unwin, London, 1961.

\item [3]
O. Neugebauer, {\it The Exact Sciences in Antiquity.}
Brown University Press, Providence, 1957.

\item [4]
J. Needham, {\it Science and Civilization in China.} Cambridge University Press,
Cambridge, 1954.

\item [5]
S. Kak, {\it The Architecture of Knowledge.} CSC, Delhi, 2004.

\item [6]
T. McEvilley, {\it The Shape of Ancient Thought: Comparative Studies in
Greek and Indian Philosophies.} Allworth Press, New York, 2002.

\item [7]
A. Schopenhauer, {\it The World as Will and Representation,} E.F.J. Payne (tr.). 
Dover, New York, 1969; page 48.

\item [8]
D. Shea and A. Troyer, {\it The Dabistan or School of Manners.}
M. Walter Dunne, Washington and London, 1901.

\item [9]
E.C. Sachau, {\it Alberuni's India.} Low Price Publications, Delhi, 1989 (1910).

\item [10] 
R.P. Kangle, {\it The Kau\d{t}il\={\i}ya Artha\'{s}\={a}stra}. Motilal
Banarsidass, 1986. See, section 1.2.11.

\item [11]
K.M. Ganguly, {\it The Mahabharata.} Munshiram Manoharlal, 1991.

\item [12]
S.C. Vidyabhusana, {\it The Ny\={a}ya S\={u}tras of Gotama}, revised and
edited by Nandalal Sinha. Motilal
Banarsidass, Delhi, 1990.

\item [13]
S.C. Vidyabhusana, {\it A History of Indian Logic}. University of Calcutta,
Calcutta, 1921;  page 49.

\item [14]
G. Feuerstein, S. Kak, D. Frawley, {\it In Search of the Cradle of Civilization:
New Light on Ancient India}. Quest Books, Wheaton, 2001 (1995).

\item [15]
B.B. Lal, {\it The Earliest Civilization of South Asia.} Aryan Books International,
New Delhi, 1997.

\item [16]
J.M. Kenoyer, {\it Ancient Cities of the Indus Valley Civilization.} Oxford 
University Press, 1998.

\item [17]
B.B. Lal, {\it The Sarasvati Flows On}. Aryan Books International, New Delhi, 2002.
S.Kak, ``Early theories on the distance to the sun.''
{\it Indian Journal of History of Science}, vol. 33, 1998, pp. 93-100.
ArXiv: physics/9804021.

\item [18]
A. Seidenberg, ``The origin of geometry.''
{\it Archive for History of Exact Sciences.} 1: 488-527, 1962.

\item [19]
A. Seidenberg, ``The origin of mathematics.''
{\it Archive for History of Exact Sciences.} 18: 301-342, 1978.

\item [20]
 S. Kak, ``Birth and early development of
Indian astronomy.'' In a book on {\it Astronomy Across Cultures:
The History of Non-Western Astronomy}, Helaine Selin (editor),
Kluwer Academic, Boston, 2000, pp. 303-340.
ArXiv: physics/0101063.

\item [21]
S. Kak, {\it The Astronomical Code of the \d{R}gveda.} Munshiram
Manoharlal, New Delhi, 2000.

\item [22]
S. Kak, ``Concepts of space, time, and consciousness in ancient India.''
In  S. Kak, {\it The Wishing Tree}, Munshiram Manoharlal, 2002; also
ArXiv: physics/9903010.

\item [23]
S. Kak, ``Physical concepts in Samkhya and Vaisesika.'' 
In {\it Life, Thought and Culture in India (from c 600 BC to c AD 300)}, edited by G.C. Pande, ICPR/Centre
for Studies in Civilizations, New Delhi, 2001, pp. 413-437;
ArXiv: physics/0310001.

\item [24]
S. Kak, ``Greek and Indian cosmology: review of early history.''
In {\it The Golden Chain}. G.C. Pande (ed.). CSC, New Delhi, 2005 (in press);
ArXiv: physics/0303001.

\item [25]
S. Kak,
``Babylonian and Indian astronomy: early connections.''
In {\it The Golden Chain}. G.C. Pande (ed.). CSC, New Delhi, 2005 (in press);
ArXiv: physics/0301078.

\item [26]
S. Kak, ``Yajnavalkya and the origins of Puranic cosmology.''
{\it Adyar Library Bulletin}, vol 65, pp. 145-156, 2001. Also in
ArXiv: physics/0101012.

\item [27]
S. Kak, ``On Aryabhata's planetary constants.'' {\it Annals Bhandarkar Oriental 
Research Institute}, vol. 84, pp. 127-133, 2003. ArXiv: physics/0110029.

\item [28]
S. Kak, ``The speed of light and Puranic cosmology.''
In {\it Computing Science in Ancient India}, T.R.N. Rao and S. Kak (eds.), USL Press,
Lafayette, 1998, Munshiram Manoharlal, New Delhi, 2000; pages 80-90;
ArXiv: physics/9804020.

\item [29]
S. Kak, ``Early theories on the distance to the sun.''
{\it Indian Journal of History of Science}, vol. 33, 93-100, 1998;
ArXiv: physics/9804021.

\item [29]
S. Kak, 
``The golden mean and the physics of aesthetics.''
ArXiv: physics/0411195.

\item [30]
S. Kak,
``Akhenaten, Surya, and the \d{R}gveda.''
In {\it The Golden Chain}. G.C. Pande (ed.). CSC, New Delhi, 2005 (in press).

\item [31]
S. Oppenheimer,
{\it The Real Eve: Modern Man's Journey Out of Africa.} Carroll \& Graf
Publishers, New York, 2003.

\item [32]
R. Lance Factor, ``What is the `logic' in Buddhist logic?''
{\it Philosophy East and West}, Volume 33, no.2, April, 1983,
pp. 183-188.

\item [33]
Rene Gu\'{e}n\'{o}n, {\it Introduction generale a l'etude
         des doctrines hindous}. Paris: 1930, pp. 226-227.

\item [34]
Diogenes Laertius, {\it Lives of the Eminent Philosophers}, Vol. 2.  R.D.
Hicks (tr.), Harvard University Press, Cambridge, 1950.

\end{description}

\end{document}